\documentclass[conference]{IEEEtran}
\IEEEoverridecommandlockouts

\usepackage[T1]{fontenc}
\usepackage{textcomp}
\usepackage{cite}

\usepackage{multicol}

\usepackage{amsmath}
\usepackage{amssymb}
\usepackage{amsthm}
\usepackage{bm}

\usepackage{caption}
\usepackage[svgnames]{xcolor}
\usepackage{listings}
\lstdefinelanguage[3]{Python}[]{Python}{morekeywords={yield,with,as}}

\newcommand{\code}[1]{\texttt{#1}}
\newcommand{\compose}{\bm{\circ}}

\newcommand{\coroutine}[2]{\left [#1; #2\right]}

\newcommand{\ConcreteTypes}{{\mathbf{K}}}
\newcommand{\Hd}{{\mathbf{h}}}
\newcommand{\Tl}{{\mathbf{t}}}
\newcommand{\Seq}[1]{{\left\langle#1\right\rangle}}
\newcommand{\Tuple}[1]{{\left ( #1 \right )}}
\newcommand{\Provided}{\textrm{provided}}

\newcommand{\Match}{\mathbf{match}}
\newcommand{\First}{\mathbf{first}}
\newcommand{\Complexity}{\mathbf{N}}

\newcommand{\csharp}{\mbox{C{\tt\#}}}
\newcommand{\cpp}{\mbox{C{\tt++}}}

\newtheorem{proposition}{Proposition}
\begin{document}

\title{Typing Composable Coroutines\\
{\footnotesize
\thanks{This work is part of the research project (RP/ESAP-02/2017) funded by Macao Polytechnic University, Macao SAR.}
}}

\author{\IEEEauthorblockN{1\textsuperscript{st} Qiqi Gu}
\IEEEauthorblockA{\textit{Faculty of Applied Sciences} \\
\textit{Macao Polytechnic University}\\
Macao SAR, China \\
qiqi.gu@mpu.edu.mo}
\and
\IEEEauthorblockN{2\textsuperscript{nd} Wei Ke}
\IEEEauthorblockA{\textit{Faculty of Applied Sciences} \\
\textit{Macao Polytechnic University}\\
Macao SAR, China \\
wke@mpu.edu.mo}
}

\maketitle

\begin{abstract}
Coroutine, as a powerful programming construct, is widely used in asynchronous applications to replace thread-based programming or the callback hell.
Using coroutines makes code more readable and maintainable, for its ability to transfer control while keeping the literal scope.
However, reasoning about coroutine behavior can be challenging without proper typing.
We propose a type notation and calculus for composing asymmetric, first-class, stackless coroutines.
Given the types of a list of coroutines, we can compute a composed type matching the collective behavior of the coroutines,
so that the input and output can be type-checked by a type system.
Our coroutine types can model the data received by or yielded from a coroutine, which be of coroutine types as well.
On top of our type calculus, we discuss its soundness and evaluation issues, then provide four application scenarios of our coroutine types.
Not only can our types be used in modern programming languages, such as Python, but also model program behaviors in OCaml and even Prolog.
\end{abstract}


\section{Introduction}

Coroutine is a generalization of function.
Regular functions always start at the beginning and exit at the end, whereas coroutines can additionally suspend execution part-way for later resumption.
A simple function type is not enough to capture the data flow during suspensions of a coroutine~\cite{anton2010towards}.

Coroutine is a powerful programming construct that provides several benefits when used in the right contexts.
For example, coroutines can replace thread-based programming or the callback hell in IoT devices. Asynchronous code is known to be challenging to implement and debug because application logic is
split between the function that initiates the request and the event handler that is invoked~\cite{levis2002mate,gay2003nesc,meijer2010reactive}.
Hence, readers have to jump around in order to understand the confusing code~\cite{brodu2015toward, madsen2017model}.
On the other hand, the most significant benefit coroutines bring is code style and simplicity~\cite{belson2019survey}.
A coroutine is a self-contained structure that encloses all variables and procedures for its complete execution, while holes (yielding and receiving points) are left for external interactions.
Thus, the program's behavior in these self-contained structures are easy for developers to reason about even if this literal scope encapsulates complex logic or business rules.
Not only in asynchronous programming, coroutines can also be used in the producer and consumer pattern with streams for communication~\cite{purely-functional}.

Coroutines can be organized as components or libraries, and further integrated into other programs.
When there is a list of coroutines, the execution of them can be made deterministic in a single threaded environment.
The processor can execute these coroutines in a first-come-first-serve order until the one being executed yields or waits for input.
Each time that the processor's attention is directed to one of the coroutines, execution is continued from the point of last abandonment.

To better understand the input and output, and the collective behavior of multiple coroutines,
we propose a type notation and calculus for coroutines.
Our type calculus is a component of a type system which gives the types of coroutines and their order, then we can calculate a composed type.
Then the type system can view this group of coroutines as a single function/coroutine for type checking purpose.

Our coroutine types are based on dependent types~\cite{bove2008dependent} that use scalars or values to more precisely describe the type of a coroutine.
Take a zip function~\cite{henriksen2021towards}, we denote its coroutine type as $\coroutine{\Seq{s^p,t^p}}{\Seq{s,t}^p}$, meaning it takes two lists of type $s$ and $t$ as parameters and returns a list of type $\Seq{s,t}$. The detail of the grammar will be explained shortly in Section~\ref{sec:syntax}.

The coroutines we support are asymmetric, first-class, and stackless.
Our syntax ignores the creation and deletion of coroutines~\cite{marlin1980coroutines}, but emphasizes their {\it resume} and {\it yield} operations~\cite{moura2009revisiting}.
The first time a coroutine is resumed, we call it is activated~\cite{de2004coroutines}.

The {\it resume} operation has form \code{resume $l$ $e$}, where $l$ is a label or variable referring to a coroutine. This command allows the caller to resume a coroutine with data $e$. From $l$'s point of view, it {\it receives} data.
The {\it yield} operation has form \code{yield $e$}. The caller of this command, being a coroutine, yields control and pops data $e$.
Therefore, each coroutine has a receiving part and a yielding part.
Given our coroutines are first-class, $e$ can be coroutines as well.

Finally, we support functions on the type level. As we show in Section~\ref{sec:applications}, $\coroutine{y^n}{y^{dec(n)}}$ is a coroutine that receives a generic list $y^n$ and yields a list with one less element $y^{dec(n)}$. However, within the type function $dec$, no {\it yield} operation is permitted. That's what we call stackless.

The remainder of this paper is organized as follows.
We first describe the syntax of our coroutine types in Section~\ref{sec:syntax}, and then the composition rules in Section~\ref{sec:rules}.
Next, Section~\ref{sec:soundness} discusses the complexity and termination of our typing rules.
In Section~\ref{sec:applications}, we provide examples of how our composable coroutine types express structures in various programming languages.
Then definitions of related concepts and projects of related work are examined in Section~\ref{sec:related-works}. Finally, we conclude in Section~\ref{sec:conclusion}.

\section{Syntax of Coroutine Types}
\label{sec:syntax}

\begin{figure}
\[
\begin{aligned}
t ::= & & & \mbox{types} \\
    & k & & \mbox{concrete types} \\
 |\ & \Seq{t,t} & & \mbox{sequences} \\
 |\ & \coroutine{t}{t} & & \mbox{coroutines of receiving and yielding types}
\end{aligned}
\]
\caption{Abstract syntax of coroutine types}
\label{fig:syntax}
\end{figure}

Fig.~\ref{fig:syntax} defines the basic syntax for our coroutine types.
A type $t$ can be a concrete type $k \in \ConcreteTypes$, a sequence $\Seq{t_1,t_2}$, or a coroutine $\coroutine{t_1}{t_2}$.
A sequence is flat, thus satisfies the associative law, $\Seq{\Seq{t_1,t_2},t_3} = \Seq{t_1,\Seq{t_2,t_3}}$, where both sides can be simplified to $\Seq{t_1,t_2,t_3}$.
A sequence of a single type $t$ is called a list of $t$, written as $t^n$, where $n$ is the length.
A list of indefinite length is denoted as $t^\ast$, while a list of zero length is denoted as $\varnothing$.
Variables, starting with lowercase letters, may also denote a type, such as $x, y, path$.
Functions that produce a valid type can be used in place of a literal type.
A concrete type $k \in \ConcreteTypes$ is a simple type, e.g., $\mathrm{Path}, \mathrm{Number}$. It doesn't consist of other types.
In places where a type $t$ is needed, a concrete type can be there.
The language is monomorphic and case-sensitive.

The first type of a coroutine is called the receiving part and the second type is called the yielding part.
The two parts are a protocol to control when to execute a coroutine.
$t^\ast$ can occur at the end of the yielding part, meaning the coroutine can be called unlimited times.
Take $\mathit{fib}:\coroutine{\Seq{\mathrm{Int}, \mathrm{Int}}}{\mathrm{Int}^\ast}$. Given two initial numbers, each time this coroutine is called, it yields the next Fibonacci number.

\section{Composition Rules}
\label{sec:rules}
In this section, we define the compose function $\compose $ that takes $\Theta$ a sequence of coroutine types, simplifies them, and returns a composition type.
The compose function runs by iterating all rules defined below, from Rule 1 to Rule 8.
The order of its coroutines in $\Theta$  should match the order when they are {\it activated} in the code. Because coroutines voluntarily yield control, if the first coroutine does not give up, other coroutines have no chance to run.
To compose the list of coroutine types, we employ a demand-driven strategy~\cite{papazoglou1984outline}.

A coroutine with empty receiving part and empty yielding part is regarded as useless and should be removed from the argument of $\compose$.
A sequence cannot contain $\varnothing$ either. Rule~\ref{remove-void} lists the composition rule regarding removing these identity elements.

\begin{equation}\label{remove-void}
\begin{split}
\compose(\Seq{t_1,\coroutine{\varnothing}{\varnothing},t_2}) &\Rightarrow \compose(\Seq{t_1,t_2}) \\
\compose(\Seq{t_1,\varnothing^\ast,t_2}) &\Rightarrow \compose(\Seq{t_1,t_2}) \\
\compose(\Seq{t_1,\varnothing, t_2}) &\Rightarrow \compose(\Seq{t_1,t_2})
\end{split}
\end{equation}

In a coroutine, the receiving part must be fully satisfied before running the yielding part.
Interleaving receiving and yielding is impossible.
Hence, variables in the receiving part can be bound and used later in the yielding part.
We require to run the receiving part first in order to model the worst scenario of a function which need to receive data before starting to yield.
As a result, yielding to itself will happen no more.

The rest of typing rules need to access auxiliary data.
Thus symbol $\vdash$ is employed. Left to $\vdash$ is the typing context
containing Pending Type $t$ and External Yields $E$.
Right to $\vdash$ is a type or a call to the compose function.
We sometimes omit unimportant items in rules, for example $(\cdot,E)$ means at this point we don't care the pending type which may or may not be $\varnothing$.

\begin{gather}
(\varnothing, E) \vdash \compose(\coroutine{s}{t}) \Rightarrow
  \coroutine{s}{\Seq{E,t}}
\label{eq:e-to-one}
\end{gather}

Rule~\ref{eq:e-to-one} is a terminal step in that the right-hand side of $\Rightarrow$ is not a call to $\compose$, meaning composition is done.
The composition result is one coroutine of which the yielding part is a sequence of $E$ and $t$.
$E$ is ahead of $t$ because $E$ is what other coroutines have already yielded.

Fig.~\ref{head-tail} defines the head and the tail of a type, where $s,t$ are general types.
We need the two functions because types in the yielding and receiving parts can be a sequence, of which the head is its first element.
When the yielding part is a sequence, elements are yielded one by one, and the yielded element becomes pending type.
When the receiving part is a sequence, the head of the sequence is the demand, driving the composition.
For all other types, the head is itself and the tail is nothing.

With the head and tail function, we are ready to detail the rules with respect to the yield operation and the resume operation.

\begin{figure}
\setlength{\belowdisplayskip}{0pt}
\setlength{\belowdisplayshortskip}{0pt}
\[
\begin{aligned}
\Hd(k) &= k \qquad & (k \in \ConcreteTypes) \\
\Hd(\coroutine{s}{t}) &= \coroutine{s}{t} \\
\Hd(\Seq{s,t}) &= \Hd(s) \\
\Tl(k) &= \varnothing & (k \in \ConcreteTypes) \\
\Tl(\coroutine{s}{t}) &= \varnothing \\
\Tl(\Seq{s,t}) &= \Seq{\Tl(s),t}
\end{aligned}
\]
\caption{Definition of head $\Hd$ and tail $\Tl$ of a type}\label{head-tail}
\end{figure}

\subsection{Yielding}

We use a function $\First$ to find the first coroutine $\theta$ in a list $\Theta$ of coroutines that matches a condition $p$, written as
$(\theta, \Theta_1, \Theta_2) = \First(\Theta, \lambda\theta.p(\theta)),$
where along with the coroutine, $\First$ also returns $\Theta_1$ all elements before $\theta$ and $\Theta_2$ all elements after $\theta$.
If $\First$ cannot find a matching element, $\theta = \Theta_2 = \varnothing$.

\begin{multline}
(\varnothing, \cdot) \vdash \compose(\Theta) \Rightarrow
  (\Hd(s), \cdot ) \vdash \compose(\Seq{\Theta_1, \coroutine{\varnothing}{\Tl(s)}, \Theta_2})
   \quad\Provided \\
(\theta,\Theta_1, \Theta_2) = \First(\Theta, \lambda \theta. \theta=\coroutine{\varnothing}{s} \wedge s \neq \varnothing)
\label{eq:yield}
\end{multline}

In list $\Theta$ we use the $\First$ function to find the first coroutine whose receiving part is $\varnothing$, and pop the head of its yielding part. The yielded type is put into the context.
When a coroutine has exhausted  its action statements, its yielding part $\Tl(s)$ would be $\varnothing$ and it's subject to deletion by Rule~\ref{remove-void}.

\begin{multline}
(\varnothing, \cdot )\vdash \compose(\Theta) \Rightarrow
  (\varnothing, \cdot) \vdash \compose(\Seq{\Theta_1, \coroutine{\varnothing}{\Tl(s)}, \coroutine{u}{v}, \Theta_2})  \\\Provided\;
(\theta,\Theta_1, \Theta_2) = \First(\Theta, \lambda \theta. \theta=\coroutine{\varnothing}{s} \wedge  \Hd(s)=\coroutine{u}{v})
\label{eq:yield-co}
\end{multline}
Being a special case of Rule~\ref{eq:yield}, Rule~\ref{eq:yield-co} here says if the yielded type is a coroutine, we instead add it after $\theta$.
In Hoare Logic, we have
$\left\{ \left| \Theta \right| = p\right\} \compose_{\ref{eq:yield-co}}(\theta) \left\{ \left| \Theta \right| = p+1 \right\}$ without considering identity removal of Rule~\ref{remove-void}.

\begin{multline}
(\varnothing, E) \vdash \compose(\Theta) \Rightarrow
  (\varnothing, \varnothing) \vdash \Seq{E,\Theta_1}\quad\Provided\\
  (\varnothing,\Theta_1, \varnothing) = \First(\Theta, \lambda \theta. \theta=\coroutine{\varnothing}{\cdot})
\label{eq:co-to-ext}
\end{multline}
In addition to Rule~\ref{eq:e-to-one}, Rule~\ref{eq:co-to-ext} is another terminal step.
If there is a call $\compose(\Theta)$ where if all coroutines in $\Theta$ cannot yield and receive, e.g., the deadlock state,
the composition result is the external yield plus all remaining coroutines.

\subsection{Resuming}

\begin{multline}
(t, \cdot) \vdash \compose(\Theta) \Rightarrow
  (\varnothing, \cdot) \vdash \compose(\Seq{\Theta_1, \coroutine{\Tl(s)}{u}[D], \Theta_2 }) \; \Provided \\
(\theta,\Theta_1, \Theta_2) = \First(\Theta,\lambda \theta. \theta=\coroutine{s}{u} \wedge \Match(t,s)=D)
\label{eq:resume}
\end{multline}
When pending type $t$ is not null, the resume operation starts. We find the first coroutine $\theta$ that can receive $t$, and resume it.
$\Match(\cdot,\cdot)$, defined in Fig.~\ref{eq:match}, checks if two types can match and returns conditions $D$.
Conditions $D$ are in the form of variable bindings as $\theta$ may contain variables. For example $\Match(Int^n,Int^5)=\{n=5\}$. If in no way can two types match, $\bot$ is returned.
The absorbing element $\bot$ joining with any condition is $\bot$.

\begin{figure}
\setlength{\belowdisplayskip}{0pt}
\setlength{\belowdisplayshortskip}{0pt}
\begin{align*}
\Match(\coroutine{s}{t}, \coroutine{u}{v}) &= \Match(s,u) \cup \Match(t,v) \\
\Match(s, \coroutine{\cdot}{\cdot}) &= \bot \\
\Match(s, t) &= \begin{cases}
 D & \mbox{if $\exists D.\Hd(t)[D] = s$}, \\
\bot & \mbox{ otherwise.}
\end{cases}
\end{align*}
\caption{Definition of $\Match$ for matching two types.}\label{eq:match}
\end{figure}


\begin{multline}
(t ,E)\vdash \compose(\Theta)  \Rightarrow (\varnothing, \Seq{E,t}) \vdash \compose(\Theta) \quad\Provided \\
(\varnothing,\Theta_1, \varnothing)=\First(\Theta, \lambda \theta. \theta=\coroutine{s}{\cdot } \wedge \Match(t,s)=\bot)
\label{eq:external}
\end{multline}
If none of the coroutines can receive the pending type $t$, we append $t$ to $E$.

\begin{multline}
(\varnothing, \cdot )\vdash \compose(\Theta)  \Rightarrow (\varnothing, \cdot ) \vdash \compose(\Seq{\Theta_1, \coroutine{\Tl(s)}{u}[D], \Theta_2} -\theta') \\\Provided\;
(\theta,\Theta_1, \Theta_2) = \First(\Theta,\lambda \theta. \theta=\coroutine{s}{u} \wedge \Hd(s)=\coroutine{\cdot}{\cdot}) \\\textrm{and}\;
(\theta', \cdot, \cdot) = \First(\Theta-\theta,\lambda \theta'. \Match(\theta', \Hd(s))=D)
\label{eq:resume-co}
\end{multline}
While Rule~\ref{eq:yield-co} yields coroutines, Rule~\ref{eq:resume-co} here permits a coroutine in $\Theta$ to receive other coroutines.
Both rules are essential for the support for first-class coroutines.
Line 2 in Rule~\ref{eq:resume-co} says in $\Theta$, we find the first coroutine $\theta$ of which head is a pattern that can receive coroutines.
Then line 3 says this pattern $\Hd(s)$ matches another coroutine $\theta'$.
A coroutine cannot receive itself, so $\Theta-\theta$ is the foremost parameter of $\First$.
In that case, we remove the received coroutine from the result. The minus sign means removing an element from a list.

Right now, with the fundamental rules introduced, we are ready to expand the syntax defined in Fig.~\ref{fig:syntax} for additional support to first-class coroutines.
The new syntax allows to label a coroutine and reference it.

\[
\begin{aligned}
t ::= & \cdots & & \\
  |\  & l: \coroutine{t}{t} & & \mbox{labeled coroutines} \\
  |\  & l^*: \coroutine{t}{t} & & \mbox{labeled infinite coroutines} \\
  |\  & l & & \mbox{references to coroutines}
\end{aligned}
\]

If we assign $\mathit{a}:\coroutine{t}{\mathrm{Int}}$, then $\coroutine{\mathrm{String}}{a}$ is equal to $\coroutine{\mathrm{String}}{\coroutine{t}{\mathrm{Int}}}$. The employment of label $l$ allows a coroutine to refer to itself and others. For example $l:\coroutine{\mathrm{S}}{(\mathrm{U},l)}$ is a coroutine that first yields U, then yields $l$, the original form of itself. We add a star ($*$) after the label to denote the same effect, which as a syntax sugar. When a starred coroutine finishes execution, it always restores itself.

\section{Soundness}
\label{sec:soundness}

If a type system is sound, it rejects all incorrect programs~\cite{maidl2014typed}.
We believe our composition rules are sound in that the rules always compute a valid composed type matching the behavior of the given coroutines, without introducing any invalid types.

The compose function $\compose $ introduced earlier tries to reduce the complexity of the given list of coroutine types.
Here we measure and compare the complexity of the input types and returned type.
By analyzing the number of types, we also conclude that the composition process may not terminate when labeled infinite coroutines are present.

At the end, we discuss the importance of evaluation order, which can affect the composition result. We also present our rationale for using the $\First$ function that always evaluates coroutine in $\Theta$ sequentially from the beginning to the end.

\subsection{Type Complexity}
The composition process generally cancels yielding and receiving types in $\Theta$, and the only remaining ones go to $E$; but some rules may increase the number of types in $\Theta$.

We adopt a case analysis approach to examine the changes in the number of types, which we call type complexity.
To begin with, we define a ranking function $\Complexity$ in Fig.~\ref{eq:complexity-function} so that the complexity of a type can be measured~\cite{cook2011proving}.

\begin{figure}
\setlength{\belowdisplayskip}{0pt}
\setlength{\belowdisplayshortskip}{0pt}
\begin{align*}
\Complexity(\varnothing) &=0 \\
\Complexity(k) &= 1 \\
\Complexity(\coroutine{s}{u}) &=\Complexity(s)+\Complexity(u)+1 \\
\Complexity(\Seq{s,u}) &=\Complexity(s)+\Complexity(u)
\end{align*}
\caption{function $\Complexity$ measures complexity of types}\label{eq:complexity-function}
\end{figure}

$\compose_{a,b}(\Theta)$ means we run Rule $a$ then Rule $b$ on the compose expression.


\begin{proposition}
When the applied rules are {\tt yield} (Rule~\ref{eq:yield}) and {\tt resume} (Rule~\ref{eq:resume}),
we assume the complexity of the yielded type is $x$ and let $E'$ and $\Theta'$ be the states after the transition,
then the following condition in Hoare Logic holds.

\begin{multline*}
\left\{ \Complexity(E)+ \Complexity(\Theta)=p \right\} E \vdash \compose_{\ref{eq:yield},\ref{eq:resume}}(\Theta) \\ \left\{p-2-2x \leqslant \Complexity(E')+ \Complexity(\Theta') \leqslant p- 2x \right\}
\end{multline*}
\end{proposition}

\begin{proof}
Assume $\theta_i$ yields, and resumes $\theta_j$. $\theta_i$ becomes $\theta_i'$; $\theta_j$ becomes  $\theta_j'$.
We have $\Complexity(\theta_i)=\Complexity(\theta_i')+x, \Complexity(\theta_j)=\Complexity(\theta_j')+x $.
After the composition, the complexity of  $\Theta$ is $p-\Complexity(\theta_i)+\Complexity(\theta_i')-\Complexity(\theta_j)+\Complexity(\theta_j') = p-2x$.
Further, Rule~\ref{remove-void} can kick in, removing $\theta_i$ or $\theta_j$ if they becomes $\coroutine{\varnothing}{\varnothing}$.
$\Complexity(E)$ does not change.
\end{proof}

\begin{proposition}
If the applied rules are {\tt yield} (Rule~\ref{eq:yield}) and {\tt external} (Rule~\ref{eq:external}), and the complexity of the yielded type $t$ is $x$, we have

\begin{multline*}
\left\{\Complexity(E)+ \Complexity(\Theta)=p \right\} E \vdash \compose_{\ref{eq:yield}, \ref{eq:external}}(\Theta)\\\left\{p-1 \leqslant \Complexity(E')+ \Complexity(\Theta') \leqslant p \right\}
\end{multline*}
\end{proposition}

\begin{proof}
Assume $\theta_i$ yields and becomes $\theta_i'$. We have $\Complexity(\theta_i)=\Complexity(\theta_i')+x$, $\Complexity(\Seq{E,t})=\Complexity(E)+x $.

\begin{align*}
  & \Complexity(E')+\Complexity(\Theta') \\
= & \Complexity(\Seq{E,t}) + \Complexity(\Theta)-\Complexity(\theta_i)+\Complexity(\theta_i') \\
= & \Complexity(E)+x + \Complexity(\Theta) -x\\
= & p
\end{align*}

If Rule~\ref{remove-void} kicks in, the complexity will further be decreased by 1.
\end{proof}

We have a few more cases of rule application. Proofs are omitted but can be made by similar techniques shown above.

\begin{proposition}
If Rule~\ref{eq:yield-co} ({\tt yield-coroutine}) is applied, we have

\begin{multline*}
\left\{ \Complexity(E)+ \Complexity(\Theta)=p \right\} E \vdash \compose_{\ref{eq:yield-co}}(\Theta) \\ \left\{p-1 \leqslant \Complexity(E')+ \Complexity(\Theta') \leqslant p \right\}
\end{multline*}
\end{proposition}

\begin{proposition}
If Rule~\ref{eq:resume-co} ({\tt consume-coroutines}) is applied, assume the complexity of consumed coroutines is $x$, we have

\begin{multline*}
\left\{ \Complexity(E)+ \Complexity(\Theta)=p \right\} E \vdash \compose_{\ref{eq:resume-co}}(\Theta) \\\left\{ p - 2x - 1  \leqslant \Complexity(E')+ \Complexity(\Theta')  \leqslant p - 2x \right\}
\end{multline*}
\end{proposition}

\begin{proposition}
If Rule~\ref{eq:co-to-ext} ({\tt coroutines-to-external}) is applied, we have
$$\left\{ \Complexity(E)+ \Complexity(\Theta)=p \right\} E \vdash \compose_{\ref{eq:co-to-ext}}(\Theta) \left\{ \Complexity(E')+ \Complexity(\Theta') = p \right\}$$
\end{proposition}

Rule~\ref{eq:co-to-ext} is a final resort of composition; this transition rule does not decrease the complexity.

Finally we discuss the impact of starred coroutines, which is capable of increasing the complexity of the system because it adds back the original form of the referred coroutine.
$a^*:\coroutine{\varnothing}{\coroutine{\mathrm{S}}{\mathrm{T}}}$ types a coroutine that yields another coroutine $\coroutine{\mathrm{S}}{\mathrm{T}}$ to $\Theta$ and also restores itself. As a result, $\Theta$ continuously expands.
$a^*:\coroutine{\varnothing}{\mathrm{S}}$ types a coroutine that yields type S to external yield $E$ and also restores itself. Therefore $E$ continuously expands.
Fig.~\ref{fig:theta-same-e-same} is a case that $\Theta$ and $E$ do not grow, yet the execution does not terminate. Here, coroutine $a$ yields coroutine $\coroutine{\mathrm{S}}{\mathrm{T}}$ which is then received by coroutine $b$. But neither $a$ or $b$ puts anything to $E$.

\begin{figure}[t]
\setlength{\belowdisplayskip}{0pt}
\setlength{\belowdisplayshortskip}{0pt}
\begin{align*}
a^*:& \coroutine{\varnothing}{\coroutine{\mathrm{S}}{\mathrm{T}}} \\
b^*:& \coroutine{\coroutine{\mathrm{S}}{\mathrm{T}}}{\varnothing}
\end{align*}
\caption{Both $\Theta$ and $E$ stay the same}\label{fig:theta-same-e-same}
\end{figure}

Overall, the complexity of the system generally reduces unless we have repeated coroutines.

\subsection{Termination of Evaluation}
Fig.~\ref{fig:theta-same-e-same} also touches the termination of our evaluation. The whole composition progress may not terminate in the presence of starred labels.
Limiting the length of $\Theta$ or $E$ does not work; a better safeguard is to limit the number of execution steps.
However if every repeating coroutine has non-empty receiving part, the system will terminate.

%

\subsection{Evaluation Order}


The way function $\First$ searches for a matching coroutine affects the end result of the composition.
Rather than finding the very first coroutine from index 0, another design choice may be to continue at the position where the last type was yielded or received. We doesn't choose this implementation because it requires to remember the last position in composition context.

Fig.~\ref{fig:starting-position} demonstrates a situation where the starting search position of Rule~\ref{eq:resume} determines the end result. After $l$ yields, if Rule~\ref{eq:resume} starts from index 0, the yielded S is received by $a$. The composition type is then $\coroutine{\mathrm{S}}{\Seq{\mathrm{T},\mathrm{U}}}$. If resuming starts from where the previous rule left, S will be received by $b$. The result becomes $\coroutine{\mathrm{S}}{\Seq{\mathrm{U},\mathrm{T}}}$.

\begin{figure}[t]
\setlength{\belowdisplayskip}{0pt}
\setlength{\belowdisplayshortskip}{0pt}
\begin{align*}
a: & \coroutine{\mathrm{S}}{\mathrm{T}}\\
l: & \coroutine{\varnothing}{\mathrm{S}}\\
b: & \coroutine{\mathrm{S}}{\mathrm{U}}
\end{align*}
\caption{The way S, yielded by $l$, resumes other coroutines determines the composition result}
\label{fig:starting-position}
\end{figure}

On the other hand, if each yielded type only has one receptor, the evaluation order does not really matter. In Fig.~\ref{fig:starting-position-not-matter}, S yielded by $l$ can only be received by $a$. T yielded by $a$ can only be received by $b$. The evaluation order is fully determined by the types.
Fig.~\ref{fig:prolog-types} in the Prolog Proof subsection demonstrates a technique where we add a tag before each type to control the data flow.

\begin{figure}[t]
\setlength{\belowdisplayskip}{0pt}
\setlength{\belowdisplayshortskip}{0pt}
\begin{align*}
a: & \coroutine{\mathrm{S}}{\mathrm{T}}\\
l: & \coroutine{\varnothing}{\mathrm{S}}\\
b: & \coroutine{\mathrm{T}}{\mathrm{U}}
\end{align*}
\caption{The end result is not affected by starting position of composition rules}
\label{fig:starting-position-not-matter}
\end{figure}

\section{Applications}
\label{sec:applications}

Types should be used in a type system and a programming language.
In this section, we demonstrate how to use our composable coroutine types in real-world type systems.
Ideally typing rules in the type system will map a language term to our type, but we do not formally specify typing rules here.
Our goal is to show the composition when types have been determined.

The first two examples use our coroutine types and their composition type to express common structures in the Python language, including \code{if-else}, \code{for} loop, \code{with}, \code{while}, and \code{try\linebreak[1]-except\linebreak[1]-finally}.
The next two examples are advanced applications, namely pattern matching and searching.

In addition to the syntax introduced in Section~\ref{sec:syntax}, we will use $t^n$ to denote a list of type $t$ of length $n$, and $\Tuple{t_1,t_2,\cdots}$ to denote tuples (product types).
Unlike sequences, tuples are not associative.

\paragraph{Mapping in Python}
\begin{lstlisting}[language={[3]python}, float=t, label=python-mapping, caption={A mapping example in Python. The type of function $f$ is $\coroutine{\mathrm{String}^n}{\Tuple{if, \mathrm{Int}}^n}$}]
def f(l):
	for s in l:
		if len(s) > 3:
			yield len(s)


for n in f(["hello", "world", "hi", "bye"]):
	print(n)
\end{lstlisting}

Listing~\ref{python-mapping} is a snippet in Python that defines a function (coroutine) \code{f} and activates it.
To find the coroutine type of \code{f}, for starters we define the inner structure $\mathit{if}: \coroutine{\mathrm{Int}}{\Seq{a:\coroutine{\mathrm{T}}{\mathrm{Int}}, b:\coroutine{y}{\varnothing}, \mathrm{X}}}$.
Then the whole function is seen as $f: \coroutine{\mathrm{String}^n}{\Tuple{\mathit{if}, \mathrm{Int}}^n}$.
This type expression says $f$ is a coroutine and when given a list of String, a list of if structures and integers will be yielded. Each integer resumes an if structure. If the integer matches certain condition which we don't know on the type level, the $\mathit{if}$ coroutine will yield an Int. In other cases X resumes $ b $ (or we say $b$ receives X), and the composition result is $\coroutine{\mathrm{Int}}{\varnothing}$. Overall, we can say the type of $\mathit{if} $ is $\coroutine{\mathrm{Int}}{\mathrm{Int}^*}$, and then $f$ is simplified to $\coroutine{\mathrm{String}^n}{\mathrm{Int}^*}$.

\begin{lstlisting}[language={[3]python}, deletekeywords={[2]{zip}}, float=t, label=python-file, caption={Open and read files, and zip lines}]
def oc1():
	yield "~/file1"

def oc2():
	yield "~/file2"

def fr(path):
	with open(path) as f:
		yield f.readlines()

def zip():
	alist = yield
	blist = yield

	i = 0
	while True:
		yield (alist[i], blist[i])
		i = i + 1

f1 = next(oc1())
f2 = next(oc2())
l1 = next(fr(f1))
l2 = next(fr(f2))

g = zip()
next(g)
print(g.send(list(l1)))
print(g.send(list(l2)))
while True:
	print(next(g))
\end{lstlisting}

\paragraph{Reading files in Python}
Listing~\ref{python-file} defines a pile of coroutines and we manually activate and schedule them on line 20 to 30. With our type expressions, it is possible to automatically compose them according to our type calculus.

The \code{oc} function can be perceived a coroutine that receives nothing but yields a value of type Path, so its type is  $ \coroutine{\varnothing}{\mathrm{Path}} $.
$ fr $ features a \code{with} structure that receives Path in order to yield a list of String and the length of the list is unknown, so its type is  $ \coroutine{\mathrm{Path}}{\mathrm{String}^*} $.
The type of $ zip $ is $ \coroutine{\Tuple{x^i, y^j}}{\Tuple{x, y}^{min(i, j)}} $.
Let $\Theta=\Seq{\mathit{oc}_1, \mathit{oc}_2, \mathit{fr}_1, \mathit{fr}_2, \mathit{zip}}$. The composition type is $ \coroutine{\varnothing}{\Tuple{\mathrm{String},\mathrm{String}}^{min(\alpha,\beta)}} $ where $ \alpha,\beta $ replace asterisks in $ \mathit{fr}_1, \mathit{fr}_2 $.

Python and many other languages support a try-except-finally structure, which can also be viewed a coroutine with type
$$\coroutine{\mathrm{Input}}{\Seq{\mathit{except}:\coroutine{\mathrm{Ex}}{\mathrm{S}}, \mathit{finally}:\coroutine{x}{\mathrm{Output}}, \mathrm{Middle}}}$$
The input of the try-except-finally coroutine is Input; the output is Output.
The coroutine yields two (sub-)coroutines before yielding Middle. If Middle is an exception ($\mathrm{Ex}$), it resumes the $\mathit{except}$ coroutine. Otherwise the $\mathit{finally}$ coroutine is resumed.
If we use a try-except-finally block in Python function \code{fr}, the coroutine type expression $\mathit{fr}$ will replace Middle.

\paragraph{Pattern Matching}
A coroutine system can model pattern matching~\cite{marlin1980coroutines}.
Listing~\ref{pattern-matching} illustrates an OCaml function \code{mem} that checks if an element is in a list.
Since line 4 has an equality check, we define the equal case as $\mathit{eq}: \coroutine{\Tuple{x, x}}{\mathrm{T}^*}$,
and the not-equal case as $\mathit{nq}: \coroutine{\Tuple{x, y}}{\varnothing}$.
Then, the type expression of \code{mem} is a composition of the $base$ case and the $rec$ case shown in Fig.~\ref{fig:pattern-matching-types}.

The $base$ coroutine shows if the length of a list is 0, F is yielded. The recursion part $rec$ is resumed if the list is longer than 0.


\begin{lstlisting}[language=ml,float=t, label=pattern-matching, caption={Check if element x is in a list}]
let rec mem x list =
	match list with
		[] -> false
		| hd::tl -> hd = x || mem x tl ;;
\end{lstlisting}

\begin{figure}[t]
\setlength{\belowdisplayskip}{0pt}
\setlength{\belowdisplayshortskip}{0pt}
\begin{align*}
base&: \coroutine{\Tuple{x, y^0}}{\mathrm{F}^*} \\
rec&: \coroutine{\Tuple{x, y^i}}{\Seq{\mathit{nq}, \mathit{eq}, \Tuple{x, y}, \Tuple{x, y^{dec(i)}}}}
\end{align*}
\caption{The types of the base case and the recursion case of function \code{mem}}
\label{fig:pattern-matching-types}
\end{figure}

Following the same order as the OCaml code, $\Theta=\Seq{\mathit{base}, \mathit{rec}_1^*, \mathit{rec}_2^*}$.
Coroutine $ \mathit{eq} $ and $ \mathit{nq} $ will be released into $\Theta$ during the execution of $\mathit{rec}$, by Rule~\ref{eq:yield-co}.
If the type of the parameters of the \code{mem} function is modeled as coroutine $\mathit{parm}: \coroutine{\varnothing}{\Seq{\mathrm{String}, \mathrm{String}^3}}$,
it is $\mathit{rec}$ that receives $\Tuple{\mathrm{String}, \mathrm{String}^3}$, and $\mathit{rec}$ goes into state $\coroutine{\varnothing}{\Seq{\mathit{nq}, \mathit{eq}, \Tuple{\mathrm{String}, \mathrm{String}}, \Tuple{\mathrm{String}, \mathrm{String}^2}}}$ by applying conditions
$D=\left\{x=\mathrm{String},\; y=\mathrm{String},\; i=3\right\}$.
The data yielded by $\mathit{rec}$ will not go back to $rec$ itself given that the receiving part is now $\varnothing$. Only when it finishes execution, because of the star, the coroutine is then restored to the original form. That's why we need $\mathit{rec}_1^*$ and $\mathit{rec}_2^*$.
Following the example input, the external yield would be $\mathrm{T}^*$, suggesting $\mathrm{String}$ is a member of $\mathrm{String}^3$.

On the flip side, if the parameter type is $\coroutine{\varnothing}{\Tuple{\mathrm{Path}, \mathrm{String}^3}}$,
$\circ(\Theta) = \Seq{\mathrm{F}^*,rec_1^*, rec_2^*, \mathit{eq}, \mathit{eq}}$. The first element $\mathrm{F}^*$ suggests $\mathrm{Path}$ is not a member of $\mathrm{String}^3$. All remaining coroutines deadlock, and according to Rule~\ref{eq:co-to-ext}, these coroutines are added to the result. A clean-up coroutine can be added to remove unused $\mathit{eq}$, but we doesn't implement here.

\paragraph{Prolog Proof}

Coroutine is also a convenient construct for implementing goal-oriented programming, such as solving Prolog queries~\cite{clocksin2003programming}.

\begin{lstlisting}[language=Prolog,float=t, label=prolog, caption={A family knowledge base in Prolog}]
child(john,sue). child(john,sam).
child(jane,sue). child(jane,sam).
child(sue,george). child(sue,gina).

male(john). male(sam). male(george).
female(sue). female(jane). female(june).

parent(Y,X) :- child(X,Y).
father(Y,X) :- child(X,Y), male(Y).

opp_sex(X,Y) :- male(X), female(Y).
opp_sex(Y,X) :- male(X), female(Y).

grand_father(X,Z) :- father(X,Y), parent(Y,Z).
\end{lstlisting}

We borrow a family knowledge base snippet found in~\cite{levesque2012thinking} and list the rules in Listing~\ref{prolog}. Given query \code{parent(X,john), female(X).}, Prolog will tell \code{X=sue}. Our compose function cannot return a Prolog solution, but it is able to prove whether a given answer is true or not.

In Fig.~\ref{fig:prolog-types}, we translate all Prolog facts to coroutines with empty yielding part. The first element of the receiving tuple is the name of the relation, which is like a tag to control where the carrying data go and come from.
A catch-all coroutine for each relation is needed, whose yielding part is No.
For each Prolog rule, such as \code{parent}, its coroutine type involves the use of variables to match how it's defined in the Prolog language.
The last yielded type of $\mathit{query}$ is Yes.
Yes and No are what we are looking at to tell if the provided answer is true or not.

\begin{figure}
\setlength{\belowdisplayskip}{0pt}
\setlength{\belowdisplayshortskip}{0pt}
\begin{align*}
\Theta=\langle \mathit{child1}:& \coroutine{\Tuple{\mathrm{Child}, \mathrm{John}, \mathrm{Sue}}}{\varnothing},\\
\mathit{child2}: & \coroutine{\Tuple{\mathrm{Child}, \mathrm{Jane}, \mathrm{Sue}}}{\varnothing},\\
\mathit{child3}: & \coroutine{\Tuple{\mathrm{Child}, \mathrm{Sue}, \mathrm{George}}}{\varnothing},\\
\mathit{child4}: & \coroutine{\Tuple{\mathrm{Child}, \mathrm{John}, \mathrm{Sam}}}{\varnothing},\\
\mathit{child5}: & \coroutine{\Tuple{\mathrm{Child}, \mathrm{Jane}, \mathrm{Sam}}}{\varnothing},\\
\mathit{child6}: & \coroutine{\Tuple{\mathrm{Child}, \mathrm{Sue}, \mathrm{Gina}}}{\varnothing},\\
\mathit{childOther}: & \coroutine{\Tuple{\mathrm{Child}, x,y}}{\mathrm{No}},\\
\mathit{female1}: & \coroutine{\Tuple{\mathrm{Female}, \mathrm{Sue}}}{\varnothing},\\
\mathit{female2}: & \coroutine{\Tuple{\mathrm{Female}, \mathrm{Jane}}}{\varnothing},\\
\mathit{female3}: & \coroutine{\Tuple{\mathrm{Female}, \mathrm{June}}}{\varnothing},\\
\mathit{femaleOther}: & \coroutine{\Tuple{\mathrm{Female}, x}}{\mathrm{No}},\\
\mathit{parent}: & \coroutine{\Tuple{\mathrm{Parent}, y, x}}{\Tuple{\mathrm{Child}, x, y}},\\
\mathit{query}: & \coroutine{x}{\Seq{\Tuple{\mathrm{Parent}, x, \mathrm{John}}, \Tuple{\mathrm{Female}, x}, \mathrm{Yes}}} \rangle
\end{align*}
\caption{The coroutine representation of the family knowledge base in Prolog}
\label{fig:prolog-types}
\end{figure}

Then if we want to verify if \code{X=sue} is a solution,
we append $\coroutine{\varnothing}{\mathrm{Sue}}$ to $\Theta$, the first type that the compose function returns is Yes.

\section{Related Work}
\label{sec:related-works}

The concept of coroutine dates back to 1960s and is proposed by Conway~\cite{conway1963design}.
There is no single precise definition of coroutine, but many researchers agree that a coroutine persists its local values between successive calls,
and the execution of a coroutine can be suspended and resumed at some later stage~\cite{marlin1980coroutines}.
A coroutine system can be categorized into three aspects, namely control-transfer mechanisms, whether coroutines are first-class objects,
and whether the control can be suspended within nested calls~\cite{moura2009revisiting}.
There are two kinds of control transfer mechanisms, namely symmetric and asymmetric.
With the symmetric one, a coroutine passes control by explicitly invoking another coroutine.
The asymmetric one provides two primitives.
A coroutine can either explicitly invoke another coroutine, in which case it establishes a parent-child relationship to the next coroutine;
or pass control to the implicit parent coroutine~\cite{anton2010towards}.
If functions within a coroutine can be suspended, we call the coroutine stackful. Otherwise it's stackless.
In this paper, the proposed type notation is for asymmetric, first-class, stackless coroutines.

Coroutines appear in various names or are branded differently in their containing languages.
Simula 67 is one of the first languages that support naming of blocks which works like a coroutine~\cite{raulefs1977semantics}.
The coroutines it supports are asymmetric, first-class, and stackless. Yet, Simula just calls such a structure an object without using the word ``coroutine''~\cite{papazoglou1984outline}.
{\cpp}20 published recently added the longed-for coroutine feature~\cite{cpp20}, which is a stackless model that requires all yield and return statements to be contained within the body of the coroutine.
Stackful coroutines are not of interest in the C/{\cpp} community, because only one out of 35 papers surveyed in~\cite{belson2019survey} implemented a stackful coroutine system in C/{\cpp}.
Golang's support for concurrent programming is based on goroutines and channels~\cite{meyerson2014go}.
Channels are typed, meaning one channel can only send and receive one type of data, but one goroutine has access to more than one channel for input and output purpose.
Goroutines are asymmetric, first-class, stackful coroutines.
If channel sending and receiving are not paired, Golang will throw runtime error ``all goroutines are asleep - deadlock''.
In the official Go compiler, goroutines are backed by threads but can be implemented in other ways as well~\cite{prell2012go}.
Our coroutines are designed to run in a single-threaded environment which comes no uncertainty of execution order.
In addition to Golang, Wren~\cite{wrenConcurrency} and Lua~\cite{de2004coroutines} also permit pausing functions within coroutines, which is a sign of stackful coroutines.
Generators generally refer to constrained asymmetric, stackless, first-class coroutines, which are found in Python~\cite{belson2019survey}, JavaScript~\cite{racordon2018coroutines}, {\csharp}~\cite{psaropoulos2019interleaving} and so on.
Fiber, found in Win32 API and {\cpp} Boost library, is a term referring to stackful coroutine~\cite{gor2018fibers}.
Cooperative threads~\cite{li2010composability}, green threads~\cite{elizarov2021kotlin}, quasi-parallel processes~\cite{wang1971coroutine} are other names of coroutines.

Although many programming languages support coroutines in one way or another, few have a type system that incorporates coroutines.
A type system is a logical system comprising a set of typing rules or judgments that assigns a type to every term in a programming language~\cite{cardelli1996type}.
In this paper, we focus on the type notations and a calculation on types returned by typing rules, ignoring the rules themselves and specific type systems.
A type system can be made ``pluggable''~\cite{bracha2004pluggable} and our coroutine types can be plugged into existing type systems and languages as well.

Typed Lua~\cite{maidl2014typed} is an attempt to give optional types to the Lua language.
The type system of Typed Lua permits type narrowing but does not express coroutine types.
Like our coroutine types, their return types can't depend on input types.
That's why we ban the interleaving of receiving and yielding part to discourage dependency between types, so that our calculus is simplified.
CorDuroy~\cite{anton2011typing} is a statically-typed language for first-class, stackful coroutines based on simply-typed lambda calculus.
CorDuroy is a complete language while our contribution is type expressions and calculus that can be plugged into existing languages.
The same author, Konrad, later implemented their idea as an OCaml library, but introducing no new syntax to OCaml~\cite{anton2010towards}.
Reference~\cite{smith2019coroutines}~discusses the impact of introducing coroutines on {\cpp} compilers with a focus on the \code{sizeof} operator.
Reference~\cite{maia2012static}~models generators in Python as $eGer(f, \Omega)$ where $f$ is the function name and $\Omega$ is the generator interface. It doesn't handle resuming through.
Flow by Facebook is a static type checker for JavaScript. It can annotate a generator function with an expression \code{Generator<\linebreak[1]Yield, Return, Next>}~\cite{flow}.
Flow largely inspired our work.
We notice many languages do not allow a return command in a coroutine, or treat the last yielded value as returned value. Hence, our coroutine type syntax doesn't include a \code{Return} part.
Nonetheless, to distinguish the yield and return operations, and go beyond, we allow the yielding and receiving part to be a sequence of diverse types.

\section{Conclusion}
\label{sec:conclusion}

Coroutine is a powerful programming construct that provides numerous benefits when used in the right contexts, including concurrency, readability and maintainability.
In this paper, we propose a case-sensitive, monomorphic type notation and calculus for composing multiple coroutine types into one type.
Given a list of coroutine types and their activation order, a composed type of these coroutines is calculated, which allows developers to call a group of coroutines as one, and a type system can detect type mismatch easier.

Albeit the exact definition of coroutine differ among computer scientists, we target coroutines that support yield and resume operations.
The coroutines that can be typed by us are first-class and stackless.
We built our system based on dependent types and the compose function models the type calculus which consists of 8 rules.
Moreover, we proved the soundness of our composition rules by showing that the number of types in the system generally decreases. We also discussed the evaluation termination problem and alternative evaluation order.
Finally applications of our type calculus are elaborated.

In the future, we are eager to integrate our coroutine type notations in a real-world type system, such as the one of Golang, to perform better type check or detect coroutine deadlock problems.
Moreover, we would like to expand the expressiveness of our notation to allow polymorphism and inheritance.
Lastly, since we are only dealing with types, type narrowing is hard.
When a fully-fledged type system is been developed, we should be able to implement branching and dependency between receiving types and yielding types.

\bibliographystyle{IEEEtran}
\bibliography{mybibliography}

\end{document}